\documentclass[aip,reprint]{revtex4-1}
\usepackage{amsmath,amssymb,amsthm}
\usepackage{bm}
\usepackage{url}
\usepackage[dvipdfmx]{graphicx,color}
\usepackage{multirow}
\usepackage{tabularx}
\usepackage[normalem]{ulem}
\usepackage[colorlinks=true,linkcolor=magenta,citecolor=magenta,breaklinks=true]{hyperref}
\usepackage[english]{babel}
\usepackage{blindtext}
\usepackage{algorithm}
\usepackage{algpseudocode}
\newcommand{\diff}{\mathrm{d}}

\newcommand{\etal}{\textit{et al}. }
\newtheorem{define}{Definition}
\newtheorem{theorem}{Theorem}

\newtheorem{problem}{Problem}

\begin{document}
\title[The lower bound of the critical connectivity]{The lower bound of the network connectivity guaranteeing in-phase synchronization}
\author{Ryosuke Yoneda}
\email{yoneda@acs.i.kyoto-u.ac.jp}
\author{Tsuyoshi Tatsukawa}
\email{tatsukawa@acs.i.kyoto-u.ac.jp}
\author{Jun-nosuke Teramae}
\email{teramae@acs.i.kyoto-u.ac.jp}
\affiliation{Graduate School of Informatics, Kyoto University, Kyoto 606-8501, Japan}

\date{\today}

\begin{abstract}
In-phase synchronization is a stable state of identical Kuramoto oscillators coupled on a network with identical positive connections, regardless of network topology. However, this fact does not mean that the networks always synchronize in-phase because other attractors besides the stable state may exist. The critical connectivity $\mu_{\mathrm{c}}$ is defined as the network connectivity above which only the in-phase state is stable for all the networks. In other words, below $\mu_{\mathrm{c}}$, one can find at least one network which has a stable state besides the in-phase sync. The best known evaluation of the value so far is $0.6828\cdots\leq\mu_{\mathrm{c}}\leq0.75$. In this paper, focusing on the twisted states of the circulant networks, we provide a method to systematically analyze the linear stability of all possible twisted states on all possible circulant networks. This method using integer programming enables us to find the densest circulant network having a stable twisted state besides the in-phase sync, which breaks a record of the lower bound of the $\mu_{\mathrm{c}}$ from $0.6828\cdots$ to $0.6838\cdots$. We confirm the validity of the theory by numerical simulations of the networks not converging to the in-phase state.
\end{abstract}
\maketitle

\begin{quotation}
Coupled phase oscillators have extensively been studied to understand synchronization being ubiquitous in nature. In-phase synchronization is always a stable state for networks of coupled identical phase oscillators regardless of network structure as far as their connection strengths are positive and identical. However, the in-phase state does not need to be a global attractor because the networks may have stable states other than the in-phase state. Previous studies have revealed that if a network is dense enough, i.e., the network's connectivity is sufficiently high, the in-phase state is a global attractor, meaning that oscillators always converge to the in-phase sync for almost all initial conditions. This result raised a natural question of how dense networks should be to ensure the global stability of the in-phase sync for networks of identical oscillators. To precisely describe the density of a network, previous studies defined the connectivity of a network as the minimum degree of the network divided by the number of nodes of the network minus one, i.e., the number of possible other nodes for each node of the network. Then they defined the critical connectivity as the smallest connectivity such that in-phase synchronization is the only stable state for any networks of identical phase oscillators as far as the connectivity of the network is greater or equal to the value. While the precise value of the critical connectivity remains unknown, many studies have refined the lower and the upper bound of the critical value. The upper bound was defined as the value above which all networks are proven not to have a stable state other than the in-phase one, and the lower bound was defined as the maximum connectivity below which at least one network is proven to have a stable state other than the in-phase one. In this paper, we develop a method to systematically analyze the stability of a class of states called twisted states of circulant networks. Using the method, we identify the highest-connectivity circulant network with a stable twisted state, thus does not converge to the in-phase sync, which provides us an improved lower bound of the critical connectivity that exceeds the existing evaluation.
\end{quotation}

\section{Introduction}
Synchronization appears in various natural and artificial phenomena and has attracted much attention in various fields. Examples of the phenomena include swinging metronomes \cite{pantaleone2002}, flashing fireflies \cite{smith1935, buck1968}, singing frogs in chorus \cite{aihara2014}, and firing of neurons \cite{cossart2003, winfree1967, Lu2016}. The coupled phase-oscillators are the widely used model of synchronization \cite{kuramoto1975}. Previous studies have revealed conditions to ensure oscillators converge to the in-phase synchronization \cite{strogatz2000, ott2008, chiba2013, daFonseca2018, Dorogovtsev2008}. However, the relationship between network structure and the tendency of synchronization has not been fully understood yet.

One of the most important questions is how synchronization depends on connectivity, or connection density, of the network \cite{watanabe1994, wiley2006, taylor2012, canale2015, ling2019, townsend2020, lu2020,kassabov2021sufficiently}. The connectivity $\mu$ of a network having $N$ nodes has been defined as the minimum degree of the nodes divided by $N-1$, the total number of other nodes. In 2012, Taylor considered networks of coupled phase-oscillators whose natural frequencies are identical and the connection among them has unit strength if it exists. For the networks, he showed that the in-phase synchronization is the only stable state if $\mu$ of a network is greater than $0.9395$\cite{taylor2012}, regardless of the structure of the network. This surprising result has attracted much attention and been refined by recent studies\cite{ling2019,lu2020,kassabov2021sufficiently}. Now it is proven that networks always synchronize if $\mu$ is greater than $0.75$\cite{kassabov2021sufficiently}. Therefore, by defining the critical connectivity $\mu_{\mathrm{c}}$ as the minimum connectivity of the networks to ensure globally stable in-phase synchronization, we can say that the best known upper bound of $\mu_{\mathrm{c}}$ is $0.75$\, while the exact value of $\mu_{\mathrm{c}}$ is not yet known.

Besides the upper bound, many studies has also revealed the lower bound of $\mu_{\mathrm{c}}$\cite{wiley2006,canale2015,townsend2020}. In particular, Townsend \etal have provided a circulant network whose connectivity is less than $0.6828\cdots$ and has a stable state other than the in-phase synchronization\cite{townsend2020}, which means that the best known lower bound of $\mu_{\mathrm{c}}$ is $0.6828\cdots$.

Previous studies, however, have used heuristic approaches rather than systematic ones to find dense networks in which competing attractors coexist with in-phase synchronization, which might have overlooked denser networks. To solve the problem, in this paper, we map the search problem to an optimization problem, namely, an integer programming problem. Following the previous study\cite{townsend2020}, we consider the circulant networks. Owing to the symmetry of the networks, we can analytically derive linear eigenvalues of the states, which enables us to formulate the optimization problem. The formulation allows us to systematically analyze a class of stable states called twisted states, which provides us an improvement on the best known lower bound from $0.6828\cdots$ to $0.6838\cdots$.

This paper is organized as follows. In Section \ref{sec:preliminary}, we introduce a model of coupled identical phase-oscillators and define the network connectivity $\mu$. In Section \ref{sec:circulant}, we consider the twisted states of the circulant networks to derive the linear eigenvalues of the states analytically. In Section \ref{sec:integer-programming}, we formulate the problem to find the densest network in which at least one twisted state is stable as an integer programming problem. We also provide a theorem yielding the rigorous solution of the optimization problem. The proof of the theorem is given in Section \ref{sec:proof}. In Section \ref{sec:bound}, we provide the maximum connectivity circulant network that has a stable twisted state, which allows us to update the lower bound of $\mu_{\mathrm{c}}$. In Section \ref{sec:numeric}, we numerically validate the results. Section \ref{sec:conclusion} gives conclusions and discussions.

\section{Preliminaries}
\label{sec:preliminary}
\subsection{Coupled identical phase-oscillators}
Identical $N$ phase-oscillators coupled with each other on a network with undirected and unit-strength interactions are defined as
\begin{align}
    \frac{\diff\theta_{i}}{\diff t} = \sum_{j=1}^{N}a_{ij}\sin(\theta_{j}-\theta_{i}),
    \label{eq:model}
\end{align}
for $i\in[N]$, where $[N] = \{1,2,\dots,N\}$.
Here, $\theta_{i}\in[0,2\pi)\simeq\mathbb{S}^{1}$ is the phase of the $i$-th oscillator and $a_{ij}$ is the $(i,j)$th-element of the $N\times N$ adjacency matrix  $A$ of the network. Since the network is undirected and unweighted, the matrix $A$ is symmetric $a_{ij}=a_{ji}\in\{0,1\}$. We also set $a_{ii} = 0$ for all $i\in[N]$ to avoid self-connection.

Note that coupled phase-oscillators have generally been defined as
\begin{align}
    \frac{\diff\theta_{i}}{\diff t} = \omega_{i} + \sum_{j=1}^{N}a_{ij}\sin(\theta_{j}-\theta_{i}),
\end{align}
for $i\in[N]$ and referred as Kuramoto model, where $\omega_{i}$ is the natural frequency of the $i$-th oscillator\cite{kuramoto1975}. Assuming that the natural frequencies are identical, $\omega_{i}=\bar{\omega}$, and rotating the whole system by $\bar{\omega}t$ recovers Eq.~\eqref{eq:model}.

\subsection{Equilibrium points and their linear stability}

Let us denote an equilibrium point of \eqref{eq:model} as $\bm{\theta}^{\ast} = (\theta_{1}^{\ast},\dots,\theta_{N}^{\ast})^{\mathsf{T}}$. Then, $\bm{\theta}^{\ast}$ satisfies
\begin{align}
    \sum_{j=1}^{N}a_{ij}\sin(\theta_{j}^{\ast}-\theta_{i}^{\ast})=0
\end{align}
for $i\in[N]$. Note that if $\bm{\theta}^{\ast}$ is an equilibrium point, $\bm{\theta}^{\ast}+c= (\theta_{1}^{\ast}+c,\dots,\theta_{N}^{\ast}+c)^{\mathsf{T}}$ is also an equilibrium point of Eq.~\eqref{eq:model} for any $c\in\mathbb{S}^{1}$ due to its rotational symmetry.

The linear stability of the equilibrium point $\bm{\theta}^{\ast}$ is determined by eigenvalues of the Jacobian matrix $J_{\bm{\theta}^{\ast}}$ whose coefficient is
\begin{align}
    \left[J_{\bm{\theta}^{\ast}}\right]_{i,j} = \left\{
    \begin{array}{ll}
        a_{ij}\cos(\theta_{j}^{\ast}-\theta_{i}^{\ast}) & i\ne j \\
        -\displaystyle\sum_{k=1}^{N}a_{ik}\cos(\theta_{k}^{\ast}-\theta_{i}^{\ast}) & i = j
    \end{array}
    \right. .
    \label{eq:jacobian}
\end{align}
All eigenvalues of the matrix are real because of its reflection symmetry, and one of them is always equal to zero due to the rotational symmetry $\bm{\theta}^{\ast}+c$. Thus, $\bm{\theta}^{\ast}$ is linearly stable if all other $N-1$ eigenvalues are negative, and it is linearly unstable if at least one of them is positive. If more than one eigenvalue is equal to zero, one needs higher-order evaluation to realize the stability analysis. In this paper, however, we only consider the linear stability of equilibrium states.

The model \eqref{eq:model} always has a trivial in-phase state, in which $\theta_{i}=0$ for all $i\in[N]$. Because $\bm{v}^{\mathsf{T}}J_{\bm{0}}\bm{v}=-\sum_{i>j}a_{ij}(v_{i}-v_{j})^{2}<0$ for any $\bm{v}=(v_{1},\dots,v_{N})^{\mathsf{T}}\in\mathbb{R}^{N}$ unless $\bm{v}=k\bm{1}$ with $k\in\mathbb{R}$, the in-phase state is always stable regardless of the network structure.

\subsection{Critical connectivity $\mu_{\mathrm{c}}$}
The \textit{connectivity} $\mu$ of a network consisting of $N$ nodes is defined as the minimum degree of the network divided by $N-1$, the maximum possible degree of the network. The devisor is $N-1$ rather than $N$ because the self-connection is not allowed. Because the degree of $i$-th oscillator is equal to the sum of the $i$-th row of the adjacency matrix $A$, the connectivity is given as
\begin{align}
    \mu = \frac{\min_{i\in[N]}\sum_{j\in[N]}a_{ij}}{N-1}.
\end{align}
The connectivity value is equal to one for the all-to-all network, while it is equal to zero for disconnected networks.

The \textit{critical connectivity} is defined as follows:
\begin{define}[Critical connectivity $\mu_{\mathrm{c}}$\cite{townsend2020}]
The critical connectivity $\mu_{\mathrm{c}}$ is the smallest value
of $\mu$ such that any network of $N$ identical phase oscillators of unit connections is globally synchronizing if $\mu\geq\mu_{\mathrm{c}}$;
otherwise, for any $\mu<\mu_{\mathrm{c}}$, at least one network having some other attractor besides the in-phase state exists.
\end{define}
The best known bound of $\mu_{\mathrm{c}}$ so far is
\begin{align}
    0.6828\cdots\leq\mu_{\mathrm{c}}\leq0.75.
\end{align}

\section{Circulant networks}
\label{sec:circulant}
Following a previous study\cite{townsend2020}, we focus on circulant networks. The circulant network is defined as a network whose adjacency matrix is a circulant matrix of the following form,
\begin{align}
A &= \left(a_{ij}\right)_{1\leq i,j\leq N}=\left(x_{j-i}\right)_{1\leq i,j\leq N}\notag\\
&=\left(
\begin{array}{ccccc}
    x_{0} & x_{1} & \dots & x_{N-2} & x_{N-1}\\
    x_{N-1} & x_{0} & x_{1} &  & x_{N-2}\\
    \vdots & x_{N-1} & x_{0} & \ddots & \vdots\\
    x_{2} &  & \ddots & \ddots & x_{1}\\
    x_{1} & x_{2} & \dots & x_{N-1} & x_{0}
\end{array}
\right),\label{eq:adjacency-mat}
\end{align}
where $x_{k} = x_{k\bmod N}$ for any $k\in\mathbb{Z}$, and $x_{0}=0$ because self-connection is not allowed now. Because $x_{i}\in\{0,1\}$ and $x_{i} = x_{N-i}$ for $i\in[N-1]$ for the undirected and unweighted networks, the structure of a circulant network is specified by the choice of $x_{1},\dots,x_{\lfloor N/2\rfloor}$ to be 0 or 1, which has $2^{\lfloor N/2\rfloor}$ possible combinations. The connectivity of the circulant network is given as
\begin{align}
    \mu=\frac{\sum_{i\in[N-1]}x_{i}}{N-1},
    \label{eq:mu}
\end{align}
because all nodes of the network share the same degree.

Townsend \etal have proven that
\begin{align}
    \bm{\theta}_{p}^{\ast} = \left(0,\frac{2\pi p}{N},\dots,\frac{2\pi p(N-1)}{N}\right)^{\mathsf{T}}
\end{align}
is an equilibrium state of the model \eqref{eq:model} on any circular networks for any $0\leq p\leq\lfloor N/2\rfloor$. Below, we refer to the state $\bm{\theta}_{p}^{\ast}$ as the $p$\textit{-twisted state}. The zero-twisted state $\bm{\theta}_{0}^{\ast}$ is the in-phase state.

Equation \eqref{eq:jacobian} gives the Jacobian matrix of the $p$-twisted state $\bm{\theta}_{p}^{\ast}$ as
\begin{align}
    \left[J_{\bm{\theta}_{p}^{\ast}}\right]_{i,j} = \left\{
    \begin{array}{ll}
        \displaystyle x_{j-i}\cos\left(\frac{2\pi p(j-i)}{N}\right) & i\ne j \\
        -\displaystyle\sum_{k=1}^{N}x_{k-i}\cos\left(\frac{2\pi p(k-i)}{N}\right) & i = j
    \end{array}
    \right. .
\end{align}
The elements of $J_{\bm{\theta}_{p}^{\ast}}$ depend only on the difference of the indices, thus setting
\begin{align}
    y_{k} = \left\{
    \begin{array}{ll}
        \displaystyle x_{k}\cos\left(\frac{2\pi pk}{N}\right) & k\in[N-1] \\
        -\displaystyle\sum_{l=1}^{N-1}x_{l}\cos\left(\frac{2\pi pl}{N}\right) & k=0
    \end{array}
    \right. ,
\end{align}
enables us to simplify the $(i,j)$-th element of the Jacobian matrix to $\left[J_{\bm{\theta}_{p}^{\ast}}\right]_{i,j} = y_{j-i}$.
Since $x_{N-k}=x_{k}$ implies $y_{N-k}=y_{k}$, $J_{\bm{\theta}_{p}^{\ast}}$ is again a symmetric circulant matrix. Using this property, we can derive its eigenvalues as
\begin{align}
    \lambda_{k}&=\sum_{l=0}^{N-1}y_{l}\cos\left(\frac{2\pi kl}{N}\right)\notag\\
    &=\sum_{l=1}^{N-1}x_{l}\cos\left(\frac{2\pi pl}{N}\right)\left[-1+\cos\left(\frac{2\pi kl}{N}\right)\right]\label{eq:ev}
\end{align}
for $k\in[N-1]$. The eigenvalue $\lambda_{0}$ is always equal to zero as we have mentioned.

\section{Integer programming}
\label{sec:integer-programming}
Consider a search problem aiming to find the densest network having a stable state besides the in-phase one. By restricting ourselves to the twisted states of the circulant networks, we can map the search problem to an optimization problem. The objective of the optimization is to maximize the connectivity $\mu$ \eqref{eq:mu} by varying $x_{i}$ under the condition that the eigenvalues $\lambda_{k}$ in \eqref{eq:ev} should be negative for all $k\in[N-1]$.

Because $x_{i}$ must be an integer, the optimization problem is expressed as a canonical form of the integer programming \cite{Conforti2014}:
\begin{problem}
    \label{prob:integer-programming}
    For $N\geq2$ and $1\leq p\leq\lfloor N/2\rfloor$,
    \begin{align}
    \begin{split}
    \textrm{maximize} & \quad\mu=\frac{1}{N-1}\bm{1}^{\mathsf{T}}\bm{x}, \\
    \textrm{subject to} & \quad \bm{x}\in\left\{0,1\right\}^{N-1},\\
    & \quad L^{(N,p)}\bm{x}<\bm{0},\\
    & \quad C^{(N)}\bm{x} = \bm{0}.
    \end{split}
    \label{eq:problem}
    \end{align}
\end{problem}
Here we defined the matrices $L^{(N,p)}\in\mathbb{R}^{(N-1)\times(N-1)}$ and $C^{(N)}\in\mathbb{R}^{(N-1)\times(N-1)}$ such that their $(k,l)$-th elements are
\begin{align}
    \left[L^{(N,p)}\right]_{k,l} = &\cos\left(\frac{2\pi pl}{N}\right)\left[-1+\cos\left(\frac{2\pi kl}{N}\right)\right],
\end{align}
and
\begin{align}
    \left[C^{(N)}\right]_{k,l} = \delta_{k,l}-\delta_{k,N-l}.
\end{align}
Because the $k$-th eigenvalue satisfies $\lambda_{k} = \left[L^{(N,p)}\bm{x}\right]_{k}$, the constraint $L^{(N,p)}\bm{x}<\bm{0}$ means that the $p$-twisted state of the $N$-body network is linearly stable. The condition $C^{(N)}\bm{x}=\bm{0}$ represents the constraint that the networks need be undirected, $x_{k} = x_{N-k}$ for $k\in[N-1]$. Intuitively, the optimization problem means that one should set as many $x_l$s as possible to $1$ while satisfying constraint conditions $L^{(N,p)}\bm{x}<\bm{0}$ and $C^{(N)}\bm{x} = \bm{0}$.

\begin{figure}[!t]
    \centering
    \includegraphics[width=8cm]{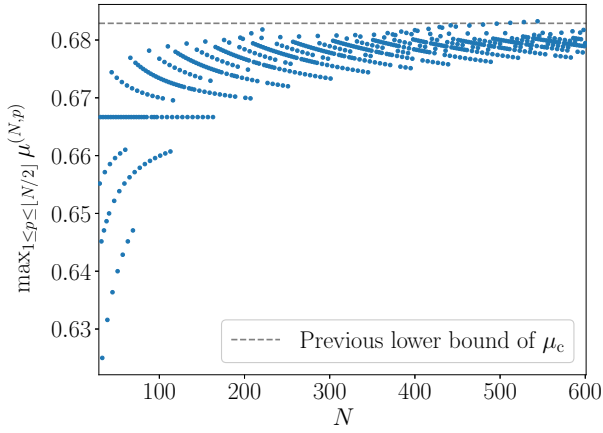}
    \caption{
    Numerical solutions $\max_{1\leq p \leq \lfloor N/2\rfloor}\mu^{(N,p)}$ of the integer programming Problem~\ref{prob:integer-programming} as a function of $N$ for $30\leq N\leq 600$. The gray dashed line shows the best known lower bound of $\mu_{\mathrm{c}}$, $0.6828\cdots$ \cite{townsend2020}. The maximum connectivity exceeds the known lower bound at $N=512$ and $544$.
    }
    \label{fig:max_connect}
\end{figure}

The conversion of the search problem into the integer programming problem enables us to systematically survey the maximum connectivity. Let $\mu^{(N,p)}$ be the solution, i.e., the maximum $\mu$, of the integer programming problem of $N$ and $p$. Figure~\ref{fig:max_connect} shows numerical solutions of $\mu^{(N,p)}$ for $30\leq N\leq 600$. We used a solver \texttt{Cbc}\cite{cbc} that can be called through the library \texttt{PuLP} in \texttt{Python} and \texttt{JuMP}\cite{DunningHuchetteLubin2017} in \texttt{Julia}\cite{Julia-2017}. (Our codes are available on \texttt{GitHub} \footnote{\url{https://github.com/yonesuke/DenseSync}}.) While further numerical computation beyond $N=600$ is intractable due to the explosion of the solution space, the search up to $N=600$ has already provided the $\mu^{(N,p)}$ that exceeds the best known lower bound at $N=512$ and $544$.

Integer programming problems are generally NP-hard \cite{Conforti2014}. However, for the specific problem, Problem \ref{prob:integer-programming}, we can obtain analytical solutions of $\mu^{(N,p)}$ for any given values of $(N,p)$. The following theorem states the result in general.
\begin{theorem}[Maximum connectivity $\mu^{(N,p)}$]
\label{th:maxmu}
For $N\geq2$ and $1\leq p\leq\lfloor N/2\rfloor$, we set $m=\gcd(N,p)$
and $\widetilde{N}=N/m$.
\begin{enumerate}
    \item For $\widetilde{N}\leq 4$, Problem~\ref{prob:integer-programming} does not have any feasible solutions.
    \item For $\widetilde{N}\geq5$, let $s_{k}$ be
    \begin{align}
        s_{k}=\sum_{l=1}^{k}\cos\left(\frac{2\pi l}{\widetilde{N}}\right)\left[-1+\cos\left(\frac{2\pi l}{\widetilde{N}}\right)\right],
    \end{align}
    and $k_{\mathrm{c}}$ be the minimum value of $k$ such that $s_{k}\geq0$.
    Then, the maximum connectivity $\mu^{(N,p)}$ is given as
    \begin{align}
        \mu^{(N,p)} = \frac{m(2k_{\mathrm{c}}-1)-3-2\left\lfloor ms_{k_{\mathrm{c}}-1}/(s_{k_{\mathrm{c}}}-s_{k_{\mathrm{c}}-1})\right\rfloor}{N-1}.     \label{eq:mu_N_p}
    \end{align}
\end{enumerate}
\end{theorem}

Note that one can easily find $k_{\mathrm{c}}$ because $s_{k}$ is a one-dimensional function of $k$. The proof of Theorem \ref{th:maxmu} is in the next section. We have observed a perfect agreement between the analytical prediction and the numerical solutions up to $N=600$ (results not shown).

\section{Proof of Theorem \ref{th:maxmu}}
\label{sec:proof}
This section gives the proof of Theorem \ref{th:maxmu}. Define a set of indices
\begin{align}
    S^{(N,p)} = \left\{l\in[N-1] \mathrel{}\middle|\mathrel{} \left\{\frac{pl}{N}\right\} \in \left[0,\frac{1}{4}\right]\cup \left[\frac{3}{4}, 1\right] \right\},
\end{align}
where $\{\alpha\}$ is the fractional part of $\alpha$. The $(k,l)$-th element of $L^{(N,p)}$ satisfies $\left[L^{(N,p)}\right]_{k,l}\leq 0$ for $l\in S^{(N,p)}$ because $\cos(2\pi pl/N)\geq 0$, whereas $\left[L^{(N,p)}\right]_{k,l}> 0$ for $l\in [N-1]\backslash S^{(N,p)}$ (please also see Fig.~\ref{fig:N60p1}). Then, using the identity for $a\in \mathbb{Z}$
\begin{align}
    \sum_{l=1}^{N-1}\cos\frac{2\pi al}{N} = 
    \left\{
        \begin{array}{ll}
            -1 & a\not\equiv 0 \bmod N\\
            N-1 & a\equiv 0\bmod N
        \end{array}
    \right. ,
\end{align}
one can show that, for $k\neq p, N-p$ and $\bm{x}\neq \bm{1}$,
\begin{align}
    &\left[L^{(N,p)}\bm{x}\right]_{k}\\
    <&\sum_{l\in S^{(N,p)}}\left[L^{(N,p)}\right]_{k,l}
    +\sum_{l\in[N-1]\backslash S^{(N,p)}}\left[L^{(N,p)}\right]_{k,l}\\
    =&\sum_{l=1}^{N-1}\left[L^{(N,p)}\right]_{k,l}\\
    =&-\sum_{l=1}^{N-1}\cos\left(\frac{2\pi pl}{N}\right)+
    \frac{1}{2}\sum_{l=1}^{N-1}\cos\left(\frac{2\pi (p+k)l}{N}\right)\\
    &+\frac{1}{2}\sum_{l=1}^{N-1}\cos\left(\frac{2\pi (p-k)l}{N}\right)\\
    =& 0,
\end{align}
which reduces the constraint of the optimization problem to
\begin{align}
    \left[L^{(N,p)}\bm{x}\right]_{p} = \sum_{l=1}^{N-1}x_{l}\left[-\cos\left(\frac{2\pi pl}{N}\right)+\cos^2\left(\frac{2\pi pl}{N}\right)\right] < 0,
    \label{eq:constraint_p}
\end{align}
because $\left[L^{(N,p)}\bm{x}\right]_{p}=\left[L^{(N,p)}\bm{x}\right]_{N-p}$. Therefore, introducing $\bm{b}^{(N,p)} = \left(b^{(N,p)}_{1},\dots,b^{(N,p)}_{N-1}\right)^{\mathsf{T}}$ where $b^{(N,p)}_{l}=-\cos\left(\frac{2\pi pl}{N}\right)+\cos^2\left(\frac{2\pi pl}{N}\right)$, we can reduce Problem~\ref{prob:integer-programming} to the equivalent problem,
\begin{problem}[Equivalent representation of Problem~\ref{prob:integer-programming}]
    \label{prob:integer-programming-equiv}
    For $N\geq2$ and $1\leq p\leq\lfloor N/2\rfloor$,
    \begin{align}
    \begin{split}
    \textrm{maximize} & \quad\mu=\frac{1}{N-1}\bm{1}^{\mathsf{T}}\bm{x}, \\
    \textrm{subject to} & \quad \bm{x}\in\left\{0,1\right\}^{N-1},\\
    & \quad \bm{b}^{(N,p) \mathsf{T}}\bm{x}<0,\\
    & \quad C^{(N)}\bm{x} = \bm{0}.
    \end{split}
    \label{eq:equiv-problem}
    \end{align}
\end{problem}
We can easily confirm that Problem~\ref{prob:integer-programming-equiv} has no feasible solutions when $\widetilde{N}\leq4$ because $b^{(N,p)}_{l}\geq 0$ in these cases. Note that one can always set $x_l=1$ for $l\in S^{(N,p)}$ because $\left[L^{(N,p)}\right]_{k,l}\leq 0$. (Remember that the objective of the optimization problem is to set as many $x_l$s as possible to $1$ with satisfying the constraints.) One can, therefore, focus only on how many additional $x_l$ of $l\in [N-1]\backslash S^{(N,p)}$ can be $1$ with satisfying the constraint condition $\bm{b}^{(N,p) \mathsf{T}}\bm{x}<0$ and $C^{(N)}\bm{x}=\bm{0}$. In the following subsections, assuming that $\widetilde{N}\geq5$, we solve Problem~\ref{prob:integer-programming-equiv} by dividing the problem into four cases: $p=1$; $p/m=1$; $p/m\ne1$ and $m=1$; $p/m\ne1$ and $m\ne 1$.

\subsection{$p=1$}
\label{subsec:p_1}
\begin{figure}[!t]
    \centering
    \includegraphics[width=8cm]{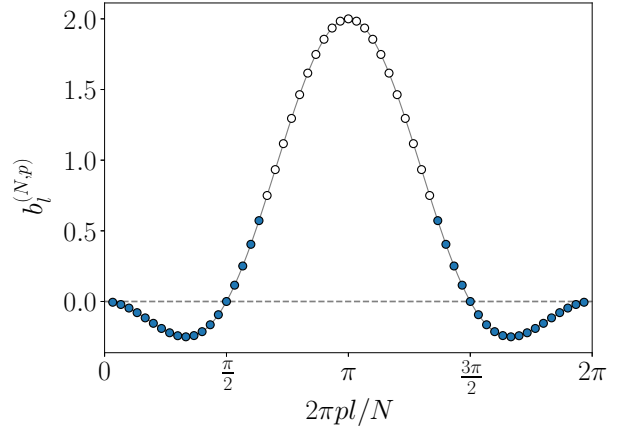}
    \caption{
        $b^{(N,p)}_{l}$ as a function of $2\pi p l/N$
        for $l\in[N-1]$ when $(N,p)=(60,1)$.
        The solid gray line is $-\cos\theta+\cos^{2}\theta$ for $\theta\in[0,2\pi]$.
        As $k_{\mathrm{c}}=20$ for $(N,p)=(60,1)$,
        $x_{k}$ can be $1$ for $k=1,2,\dots,19,41,42,\dots,59$ (filled circles) whereas other $x_{k}$s should be zero (empty circles). Note that both ends, $2\pi p l/N=0, 2\pi$, are out of the domain.
    }
    \label{fig:N60p1}
\end{figure}

Denote the cumulative sum of $b^{(N,p)}_{l}$ as
\begin{align}
    s_{k} = \sum_{l=1}^{k}b^{(N,p)}_{l}.
\end{align}
Because the function $-\cos\theta+\cos^{2}\theta$ (see solid line of Fig.~\ref{fig:N60p1} as an example) is symmetric around $\theta=\pi$ and monotonically increases from zero for $\theta\in[\pi/2,\pi]$, $s_{k}$ takes its minimum negative value at $l=\lfloor N/4 \rfloor$, i.e., when $2\pi p l/N$ is just below $\pi/2$, and monotonically increases up to $l=\lfloor N/2\rfloor$. Therefore, in order to set as many $x_{k}$ to $1$ as possible while keeping the condition $\bm{b}^{(N,p) \mathsf{T}}\bm{x}<0$, one can set $x_{k}=1$ for $k=1,\dots,k_{\mathrm{c}}-1$, and $N-k_{\mathrm{c}}+1,\dots,N-1$ due to the symmetry constraint $C^{(N)}\bm{x}=\bm{0}$, where $k_{\mathrm{c}}$ is the smallest value of $k$ such that $s_{k}\geq0$. Other $x_{k}$s of $k\in\left\{k_{\mathrm{c}},\dots,N-k_{\mathrm{c}}\right\}$ must be zero. Thus, the maximum number of $x_{k}$ that can be $1$ is $2(k_{\mathrm{c}}-1)$, which means that the maximum connectivity is
\begin{align}
    \mu^{(N,1)}=\frac{\sum_{k\in[N-1]}x^{\ast}_{k}}{N-1}=\frac{2(k_{\mathrm{c}}-1)}{N-1}.
\end{align}
This expression agrees with Eq.~\eqref{eq:mu_N_p} for the case of $p=1$ because $\lfloor s_{k_{\mathrm{c}-1}}/(s_{k_{\mathrm{c}}}-s_{k_{\mathrm{c}-1}})\rfloor=-1$. Figure \ref{fig:N60p1} shows $b^{(60,1)}_{l}$ as a function of $2\pi p l/N$, as an example. Because $s_{19}=-0.6972\cdots<0$ while $s_{20}=0.0527\cdots\geq0$, $k_{\mathrm{c}}=20$, which provides $\mu^{(60,1)}=\frac{2\cdot 19}{60-1}=0.6440\cdots$.

\subsection{$p=m$}
\label{subsec:p_m}
\begin{figure*}[thbp]
    \centering
    \includegraphics[width=\textwidth]{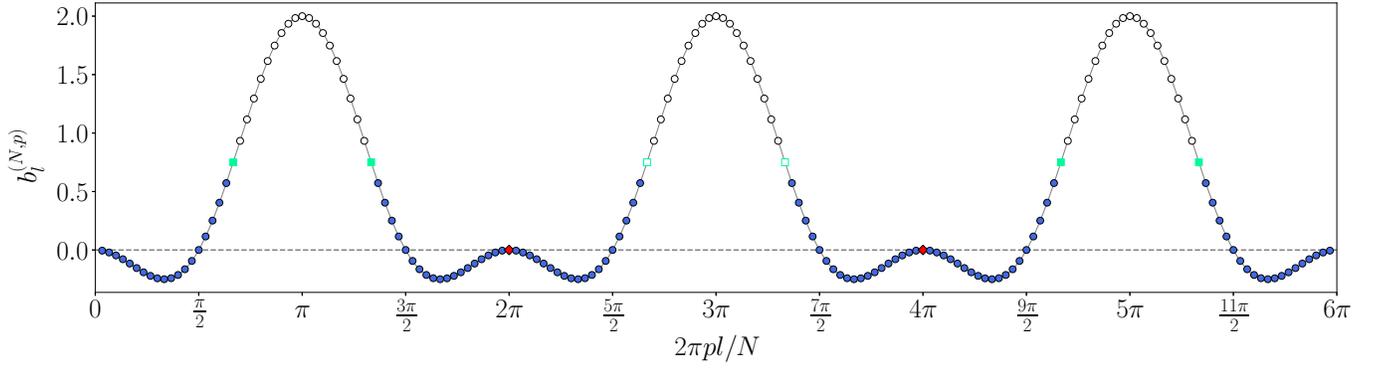}
    \caption{
        $b^{(N,p)}_{l}$ as a function of $2\pi p l/N$ for $l\in[N-1]$ when $(N,p)=(180,3)$. The solid gray line is $-\cos\theta+\cos^{2}\theta$ for $\theta\in[0,6\pi]$. $2(k_{\mathrm{c}}-1)$ $x_l$s on each $I_n, n=1,2,\dots ,m$ (blue filled circles) and $m-1$ $x_l$s on $I$ (red filled diamonds) can be $1$. Additionally, $2\left(\left\lceil\left(-ms_{k_{\mathrm{c}}-1}\right)/(s_{k_{\mathrm{c}}}-s_{k_{\mathrm{c}}-1})-1\right\rceil\right)$ (green filled squares) of $2m$ $x_l$s (green empty and filled squares) can be $1$. Note that both ends, $2\pi p l/N=0, 6\pi$, are out of the domain.
    }
    \label{fig:N180p3}
\end{figure*}

Define an integer $\widetilde{N}=N/m$ (remember that $m=\gcd(N,p)$) and divide the index domain $[N-1]$ of $x_l$s into $m+1$ disjoints subsets; $[N-1]=I+I_1+\cdots +I_m$, where $I=\left\{\widetilde{N}, 2\widetilde{N},\dots, (m-1)\widetilde{N}\right\}$ and $I_n=\left\{(n-1)\widetilde{N}+1,\dots ,n\widetilde{N}-1\right\}$. Because the function $-\cos\theta+\cos^{2}\theta$ is $2\pi$-periodic (see solid line of Fig.~\ref{fig:N180p3} as an example), $\bm{b}^{(N,p)}=\bm{b}^{(\widetilde{N},1)}$ on each $I_{n}$. Figure \ref{fig:N180p3} shows the case of $(N,m)= (180,3) $ as an example. Thus, following discussion of the previous subsection, one can set $2(k_{\mathrm{c}}-1)$ $x_l$s to $1$ on each $I_n$ with keeping $\bm{b}^{(N,p) \mathsf{T}}\bm{x}<0$ (blue filled circles in Fig.~\ref{fig:N180p3}). One can also set all $x_l$ to $1$ for $l\in I$ because $\left[b^{(N,p)}\right]_{l}=0$ on the subsets (red filled diamonds in Fig.~\ref{fig:N180p3}).

So far, the value of $\bm{b}^{(N,p) \mathsf{T}}\bm{x}$ is equal to $2ms_{k_{\mathrm{c}}-1}$ that is still negative. This implies that the possibility of additional $x_l$s being $1$ still remains. The lowest value of $\left[b^{(N,p)}\right]_{l}$ in remaining, i.e. indices for which $x_l$ has not set to $1$ yet, is $b^{(\widetilde{N},1)}_{k_{\mathrm{c}}}$ that equals $s_{k_{\mathrm{c}}}-s_{k_{\mathrm{c}}-1}$. There are $2m$ such $l$s in the domain due to periodicity and the symmetry of $\left[b^{(N,p)}\right]_{l}$ (green empty and filled squares in Fig.~\ref{fig:N180p3}). To set as many additional $x_l$s to $1$ as possible, one should use these $l$s. Therefore, one can set a maximum of
\begin{align}
    2\left(\left\lceil\frac{-2ms_{k_{\mathrm{c}}-1}}{2(s_{k_{\mathrm{c}}}-s_{k_{\mathrm{c}}-1})}-1\right\rceil\right)
    \label{eq:additional-one}
\end{align}
additional $x_l$s to $1$ (pink filled circles in Fig.~\ref{fig:N180p3}). The factor 2 being at the front and in the divisor of Eq.~\eqref{eq:additional-one} appears because one has to simultaneously set $x_l$ and $x_{N-l}$ to $1$ to keep the symmetry condition $C^{(N)}\bm{x}=\bm{0}$. Note that, as far as one keeps the numbers and conditions, one can choose any combination of $l$s from the $2m$ $l$s.

Putting the above results together, we obtain that
\begin{align}
    \mu^{(N,p)}=\frac{2m(k_{\mathrm{c}}-1)+m-1-2\left(\left\lfloor \frac{ms_{k_{\mathrm{c}}-1}}{s_{k_{\mathrm{c}}}-s_{k_{\mathrm{c}}-1}}\right\rfloor+1\right)}{N-1},
\end{align}
which agrees with Eq.~\eqref{eq:mu_N_p} of the theorem. Here we use the identity $\lceil\alpha\rceil=-\lfloor-\alpha\rfloor$ to derive the above result.

\subsection{$p\ne m,m=1$}
\label{subsec:m_1}
Because $m=\gcd(N, p)=1$, we have
\begin{align}
    \{1,2,\dots,N-1\}=\{p,2p,\dots,(N-1)p\} \pmod{N},
\end{align}
which means that $\left\{\left[\bm{b}^{(N,p)}\right]_l\right\}$ is equal to $\left\{\left[\bm{b}^{(N,1)}\right]_l\right\}$ as a set. Thus, we obtain $\mu^{(N,p)}=\mu^{(N,1)}$, and are able to reduce this case to the case of subsection \ref{subsec:p_1}.

\subsection{$p\ne m,m\ne1$}
\label{subsec:m_geq_1}
Using the same argument as before, one can see that $\left\{\left[\bm{b}^{(N,p)}\right]_l\right\}=\left\{\left[\bm{b}^{(m\widetilde{N}, m)}\right]_l\right\}$ as a set, which results in $\mu^{(N,p)}=\mu^{(m\widetilde{N},m)}$. Thus, this case is reduced to the case of subsection \ref{subsec:p_m}.

Putting all cases of subsections \ref{subsec:p_1}--\ref{subsec:m_geq_1} together, we arrive Theorem \ref{th:maxmu}.

\section{The supremum of $\mu^{(N,p)}$}
\label{sec:bound}
In this section, we derive the supremum $\overline{\mu}$ of $\mu^{(N,p)}$ defined as
\begin{align}
    \overline{\mu} :=\sup\left\{\mu^{(N,p)} \mathrel{}\middle|\mathrel{} 1\leq p\leq \lfloor N/2\rfloor, N\geq2\right\},
    \label{eq:mu_bar}
\end{align}
which leads to improvement of the lower bound of the critical connectivity $\mu_{\mathrm{c}}$.

\begin{figure}[!t]
    \centering
    \includegraphics[width=8cm]{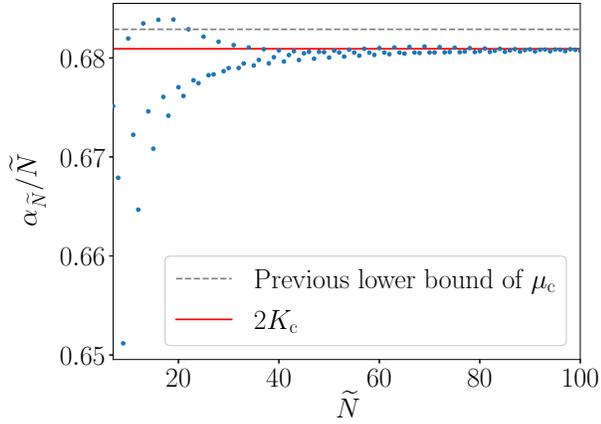}
    \caption{
        $\alpha_{\widetilde{N}}/\widetilde{N}$ for $5\leq\widetilde{N}\leq100$ (blue circles). The gray dashed line is the best known lower bound of the critical connectivity $\mu_{\mathrm{c}}$. The solid red line represents $2K_{\mathrm{c}}$.
    }
    \label{fig:sup_m}
\end{figure}

From the proof of Theorem~\ref{th:maxmu}, we have that
\begin{align}
    \overline{\mu}=\sup\left\{\mu^{(m\widetilde{N},m)} \mathrel{}\middle|\mathrel{} m\geq 1,\widetilde{N}\geq5\right\}.
    \label{eq:sup_mu_mn}
\end{align}
Then, because 
\begin{align}
    \frac{\alpha_{\widetilde{N}} m-3}{\widetilde{N}m-1}\leq\mu^{(m\widetilde{N},m)} \leq \frac{\alpha_{\widetilde{N}} m-2}{\widetilde{N}m-1},
\end{align}
where
\begin{align}
    \alpha_{\widetilde{N}}=2k_{\mathrm{c}}-1-2\frac{s_{k_{\mathrm{c}}-1}}{s_{k_{\mathrm{c}}}-s_{k_{\mathrm{c}}-1}},
\end{align}
we have 
\begin{align}
\mu^{(m\widetilde{N},m)}\leq\lim_{m\to\infty}\mu^{(m\widetilde{N},m)}=\frac{\alpha_{\widetilde{N}}}{\widetilde{N}}
\end{align}
for $m\geq 1,\widetilde{N}\geq5$ by the squeeze theorem. Here we used
\begin{align}
    \frac{\alpha_{\widetilde{N}} m-2}{\widetilde{N}m-1}
    \leq \lim_{m\to\infty}\frac{\alpha_{\widetilde{N}} m-2}{\widetilde{N}m-1} = \frac{\alpha_{\widetilde{N}}}{\widetilde{N}}
\end{align}
that follows from $k_{\mathrm{c}}\leq N/2$ and $-1\leq s_{k_{\mathrm{c}}-1}/(s_{k_{\mathrm{c}}}-s_{k_{\mathrm{c}}-1})<0$. Then there holds,
\begin{align}
    \overline{\mu}=\sup\left\{\frac{\alpha_{\widetilde{N}}}{\widetilde{N}} \mathrel{}\middle|\mathrel{} \widetilde{N}\geq5\right\}.
    \label{eq:sup_mu_n}
\end{align}
Figure \ref{fig:sup_m} shows $\alpha_{\widetilde{N}}/\widetilde{N}$ for $5\leq\widetilde{N}\leq100$.

Now let us derive the maximum of $\alpha_{\widetilde{N}}/\widetilde{N}$. We first obtain the inequality
\begin{align}
\frac{\alpha_{\widetilde{N}}}{\widetilde{N}}\leq2K_{\mathrm{c}}+\frac{2}{\widetilde{N}}+\frac{4\pi}{3\widetilde{N}^{2}}
\label{eq:alpha_ineq}
\end{align}
from
\begin{align}
    \frac{k_{\mathrm{c}}}{\widetilde{N}}\leq K_{\mathrm{c}}+\frac{1}{2\widetilde{N}}+\frac{2\pi}{3\widetilde{N}^{2}},
\end{align}
where
\begin{align}
2K_{\mathrm{c}}:=\lim_{\widetilde{N}\to\infty}\frac{\alpha_{\widetilde{N}}}{\widetilde{N}}=2\lim_{\widetilde{N}\to\infty}\frac{k_{\mathrm{c}}}{\widetilde{N}}=2\cdot 0.34046\cdots.
\end{align}
The proof of Eq.~\eqref{eq:alpha_ineq} and the derivation of the value of $ K_{\mathrm{c}}$ are given in Appendix~\ref{sec:bound_kc}. Then, from the above inequality, we have $\alpha_{\widetilde{N}}/\widetilde{N}\leq 0.683$ for $\widetilde{N}\geq1001$.

For $\widetilde{N}\leq 1000$, as Fig.~\ref{fig:sup_m} shows, some $\alpha_{\widetilde{N}}/\widetilde{N}$ exceed $0.683$. Calculating these values, we can find that the maximum is given by $\widetilde{N}=19$. Combining this with the result of the previous paragraph leads to the theorem, which sets a new lower bound of the critical connectivity exceeding the previous one:
\begin{theorem}[Supremum value of $\mu^{(N,p)}$]
    \label{th:supmu}
    \begin{align}
        \overline{\mu} =& \frac{11}{19}-\frac{2}{19}\frac{\displaystyle\sum_{l=1}^{5}\left[-\cos\left(\frac{2\pi l}{19}\right)+\cos^{2}\left(\frac{2\pi l}{19}\right)\right]}{\displaystyle-\cos\left(\frac{12\pi}{19}\right)+\cos^{2}\left(\frac{12\pi}{19}\right)}\\
        =&0.683875\cdots.
    \end{align}
\end{theorem}

The above discussion shows that the densest circulant network having a competing stable state besides the in-phase synchronization is given at $m\to\infty$ when $(N,p)=(19m,m)$.
In other words, when we increase network connectivity, the network that most persistently keeps a stable twisted state is the infinitary large network of $19m$ nodes ($m\to\infty$), and the most persistent twisted state is the $m$-twisted state.
We summarize an explicit construction of the adjacency matrix of the dense $19m$-node circulant network as Algorithm~\ref{alg1}.
As the limit of $m\to \infty$, the output of the algorithm converges to the adjacency matrix of the densest circulant network that delivers the new bound $\bar{\mu}$ along with the stable $m$-twisted state.
Whether the series of the $19m$-node network has some specific topological features remains an open question.

\begin{algorithm}[H]
    \caption{
        An explicit construction of the adjacency matrix of the dense circulant network having the stable $m$-twisted state.
        Here, $\bm{x}[i]$ is the $i$-th element of $\bm{x}$, $A[i,j]$ is the $(i,j)$-th element of $A$,
        and $b_{l}=-\cos(2\pi l/19)+\cos^{2}(2\pi l/19)$.
    }
    \label{alg1}
    \begin{algorithmic}[1]
    \Require $m$
    \State $\bm{x}\in\{0,1\}^{19m} \gets \bm{0}$
    \State $A\in\{0,1\}^{19m\times 19m}$ (Adjacency matrix)
    \State ExtraAllowance $\gets 2\left\lceil -m\sum_{l=1}^{5}b_{l}/b_{6}-1\right\rceil$
    \State $c \gets 0$
    \For{$k \gets 0$ to $m-1$}
        \For{$i \gets 1$ to $5$}
        \State $\bm{x}[19k+i],\bm{x}[19(k+1)-i] \gets 1$
        \EndFor
        \If{$k\geq1$}
        \State $\bm{x}[19k] \gets 1$
        \EndIf
    \EndFor
    \For{$k \gets 0$ to $m-1$}
    \State $\bm{x}[19k+6],\bm{x}[19(m-k)-6] \gets 1$
    \State $c \gets c+2$
    \If{$c > $ ExtraAllowance}
        \State break
    \EndIf
    \State $\bm{x}[19k+13],\bm{x}[19(m-k)-13] \gets 1$
    \State $c \gets c+2$
    \If{$c > $ ExtraAllowance}
        \State break
    \EndIf
    \EndFor
    \For{$i,j \gets 1$ to $19m$}
        \State $A[i,j] \gets \bm{x}[i-j \bmod 19m]$
    \EndFor
    \State \Return $A$ (Resulting adjacency matrix)
    \end{algorithmic}
\end{algorithm}

\section{Numerical Simulations}
\label{sec:numeric}
To validate Theorem~\ref{th:maxmu}, we numerically integrate the model~\eqref{eq:model} for $(N,p)=(1900,100)$, as an example, using the fourth-order Runge--Kutta algorithm with a time step of $\delta t=10^{-3}$. Thus $m=\gcd(N, p)=100$, $\widetilde{N}=19$, and the maximum connectivity of the network is
\begin{align}
    \mu^{(1900,100)}=\frac{1297}{1899}=0.682991\cdots,
\end{align}
which is greater than the previously reported value of the lower bound.

We set initial phases as $\bm{\theta}(0)=\bm{\theta}_{p}^{\ast}+\overline{\bm{\varepsilon}}$ to see the stability of the $p$-twisted state $\bm{\theta}_{p}^{\ast}$, where $\bm{\theta}(t)=(\theta_{1}(t),\dots,\theta_{N}(t))^{\mathsf{T}}$ and $\overline{\bm{\varepsilon}}$ is a small initial perturbation. Remember that the $p$-twisted state is the most stable twisted state now because $p=m$. The initial perturbation $\overline{\bm{\varepsilon}}$ is prepared as follows: We first draw an $N$-dimensional Gaussian random variable $\bm{\varepsilon}=(\varepsilon_{1},\dots,\varepsilon_{N})^{\mathsf{T}}$ with $\varepsilon_{i}\sim\mathcal{N}(0,\sigma^{2}/N)$ and $\sigma=0.1\pi$ and then set $\overline{\bm{\varepsilon}}=\bm{\varepsilon}-\varepsilon_{0}$, where $\varepsilon_{0}=(\sum_{i=1}^{N}\varepsilon_{i})/N$, to ensure $\sum_{i=1}^{N}\overline{\varepsilon}_{i}=0$, which is indispensable for the stability analysis because without the condition $\bm{\theta}(t)$ never converges to $\bm{\theta}_{p}^{\ast}$ due to the rotational symmetry of the model.

\begin{figure}[!t]
    \centering
    \includegraphics[width=8cm]{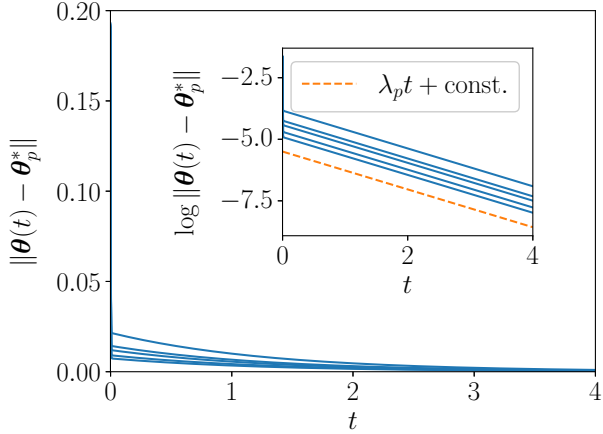}
    \caption{
        Temporal developments of $\|\bm{\theta}(t)-\bm{\theta}^{\ast}_{p}\|$ for five different initial conditions. The inset shows the semi-log plot of them. The orange dashed line in the inset represents exponential decay with the exponent $\lambda_{p}$.
    }
    \label{fig:ode}
\end{figure}

Figure~\ref{fig:ode} shows the results of the numerical simulation for realizations of the initial perturbation. To measure the distance between $\bm{\theta}(t)$ and $\bm{\theta}_{p}^{\ast}$ on the $2\pi$-periodic space, we defined a quasinorm
\begin{align}
    &\|\bm{\varphi}\| = \sqrt{\sum_{i=1}^{N}\mathrm{d}(\varphi_{i})^{2}},\\
    &\mathrm{d}(\varphi) = 
    \begin{cases}
        \varphi\bmod 2\pi & 0\leq\left(\varphi\bmod 2\pi\right) < \pi\\
        2\pi - (\varphi\bmod 2\pi) & \pi\leq\left(\varphi\bmod 2\pi\right) < 2\pi
    \end{cases} .
\end{align}
We see that the distance $\|\bm{\theta}(t)-\bm{\theta}^{\ast}_{p}\|$ monotonically decreases to zero regardless of the initial conditions, revealing that the $p$-twisted state, besides the trivial in-phase state, is stable on the dense network whose connectivity exceeds the previous lower bound.

The inset of Fig.~\ref{fig:ode} shows the developments of $\log\|\bm{\theta}(t)-\bm{\theta}^{\ast}_{p}\|$. As indicated by our analysis, the distance exponentially decreases to zero with the exponent of $\lambda_{p}=\left[L^{(N,p)}\bm{x}^{\ast}\right]_{p}$, where $\bm{x}^{\ast}$ is the binary vector specified in Sec.~\ref{sec:proof} to achieve the maximum connectivity of the network.

\section{Discussion}
\label{sec:conclusion}
In this paper, we searched for the densest networks of identical phase oscillators that have at least one attractor besides the trivial in-phase state. Focusing on the twisted states of the circulant networks, we replaced the search problem with an optimization problem, an integer programming problem, which enables us to systematically study the stability of all twisted states on all possible circulant networks. The rigorous solution of the optimization problem provides us a new record of the network connectivity $0.6838\cdots$ such that a twisted state remains stable in a dense network, in other words, the record-breaking lower bound of the critical connectivity $\mu_{\mathrm{c}}$.

Many open questions remain about the critical connectivity. While this study revealed the stability of all twisted states of all circulant networks, it remains unclear whether circulant networks have other stable states besides the twisted states. It also remains unknown whether some dense networks not included in the circulant networks have stable states that break a record of the lower bound of the critical connectivity. One may be required stability analysis beyond the linear region to answer these questions. The problem of determining the upper bound of the critical connectivity also remains open as another essential subject. Because the network model of coupled identical phase oscillators can be written as a gradient system using a potential function, geometric approaches, utilizing the Morse theory\cite{matsumoto2002} for instance, may be helpful to approach the problem.

\acknowledgments
R. Y. acknowledges the support of Iwadare Scholarship from Iwadare Scholarship Foundation.
This work was partially supported by JSPS KAKENHI Grant Number JP16H01719.

\section*{Data availability}
The data that support the findings of this study are available from the corresponding author upon reasonable request.

\appendix

\section{Upper bound of $k_{\mathrm{c}}$}
\label{sec:bound_kc}
In this appendix, we prove that
\begin{align}
    \frac{k_{\mathrm{c}}}{\widetilde{N}}\leq K_{\mathrm{c}}+\frac{1}{2\widetilde{N}}+\frac{2\pi}{3\widetilde{N}^{2}},
\end{align}
where $K_{\mathrm{c}}$ is the limit of $k_{\mathrm{c}}/\widetilde{N}$.

We first derive $K_{\mathrm{c}}$. Setting $y=l/\widetilde{N}$ gives the continuum limit $\widetilde{N}\to\infty$ of $s_{k}/\widetilde{N}$ as
\begin{align}
    &t(x) = \int_{0}^{x}b(y)\diff y,\\
    &b(y) = -\cos2\pi y+\cos^{2}2\pi y.
\end{align}
Then $K_{\mathrm{c}}$ is given as the solution of the self-consistent equation;
\begin{align}
    8\pi t(K_{\mathrm{c}})=4\pi K_{\mathrm{c}}+\sin 4\pi K_{\mathrm{c}}-4\sin 2\pi K_{\mathrm{c}}=0.
\end{align}
Conventional search algorithms such as the binary search or the Newton--Raphson method give us an approximate value of $K_{\mathrm{c}}$ as $0.34046\cdots$.

To see the difference between $K_{\mathrm{c}}$ and $k_{\mathrm{c}}/\widetilde{N}$,
we calculate $s_{k}/\widetilde{N}$ as an equation deviated from $t(k/\widetilde{N})$. In the following, we restrict the range of $k$ to $1/4\leq k/\widetilde{N}\leq 1/2$ to focus on the value of $s_{k}/\widetilde{N}$ around $k=k_{\mathrm{c}}$. From a trigonometric identity
\begin{align}
    \sum_{l=0}^{k}\cos l\theta=\frac{\sin(k\theta)}{2\tan(\theta/2)}+\cos^{2}\left(\frac{k\theta}{2}\right),
\end{align}
we rewrite $s_{k}/\widetilde{N}$ as
\begin{align}
    \frac{s_{k}}{\widetilde{N}}=&\frac{1}{2}\frac{k}{\widetilde{N}}-\frac{1}{2\pi}\frac{\pi/\widetilde{N}}{\tan[\pi/\widetilde{N}]}\sin\left(2\pi\frac{k}{\widetilde{N}}\right)\\
    &+\frac{1}{8\pi}\frac{2\pi/\widetilde{N}}{\tan[2\pi/\widetilde{N}]}\sin\left(4\pi\frac{k}{\widetilde{N}}\right)\\
    &+\frac{1}{2\widetilde{N}}b\left(\frac{k}{\widetilde{N}}\right).
\end{align}
Using an inequality
\begin{align}
    \tan x \geq x + \frac{x^{3}}{3} \quad x\geq 0,
\end{align}
we have
\begin{align}
    \frac{s_{k}}{\widetilde{N}} > &\frac{1}{2}\frac{k}{\widetilde{N}}-\frac{1}{2\pi}\sin\left(2\pi\frac{k}{\widetilde{N}}\right)+\frac{1}{8\pi}\sin\left(4\pi\frac{k}{\widetilde{N}}\right)\\
    &+\frac{1}{2\widetilde{N}}b\left(\frac{k}{\widetilde{N}}\right)-\frac{\pi}{3N^{2}}+\frac{5\pi^{3}}{18 N^{4}}\\
    >& t\left(\frac{k}{\widetilde{N}}\right)+\frac{1}{2\widetilde{N}}b\left(\frac{k}{\widetilde{N}}\right)-\frac{\pi}{3N^{2}}.
\end{align}

\begin{figure}
    \centering
    \includegraphics[width=8cm]{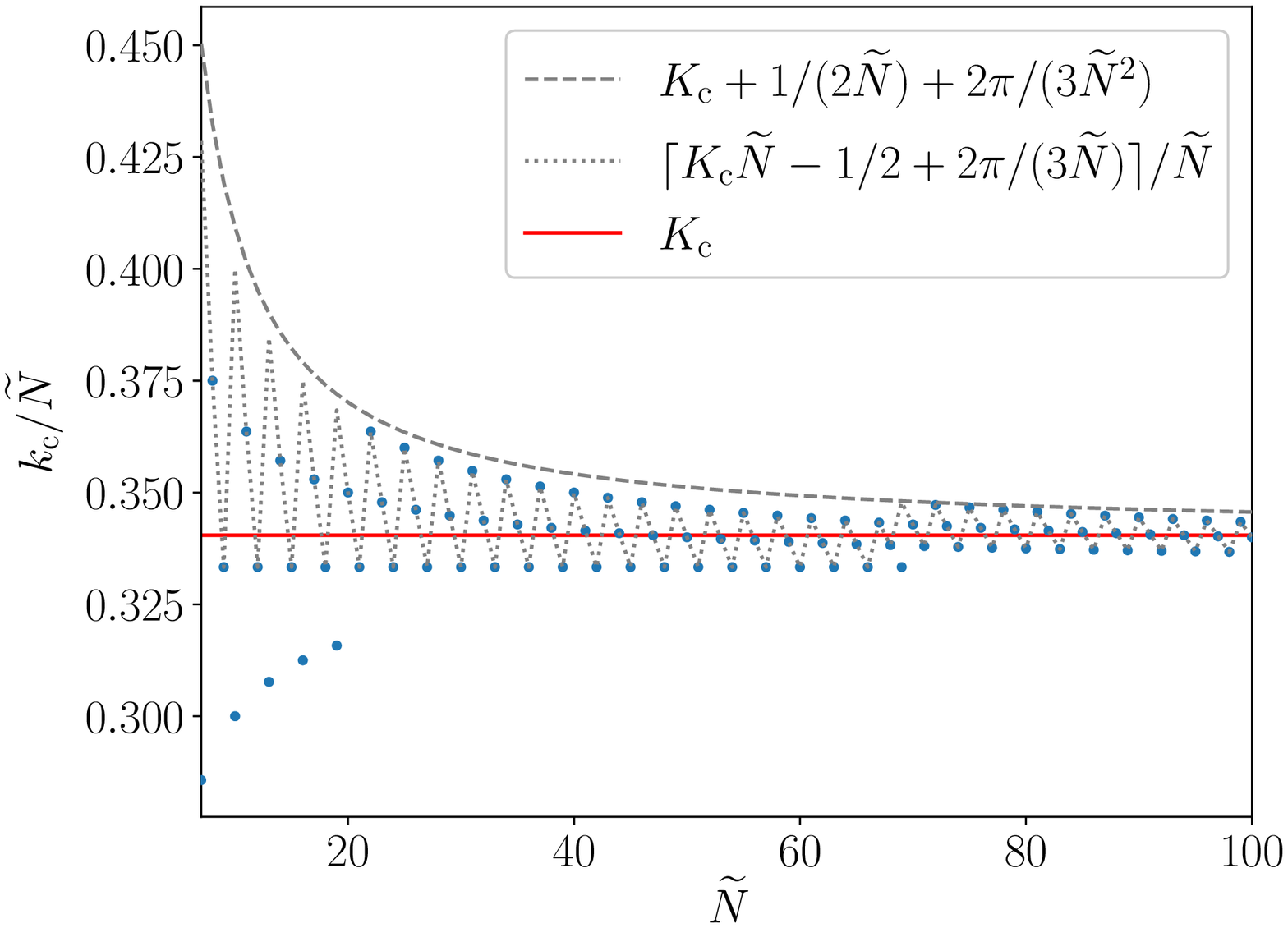}
    \caption{
        $k_{\mathrm{c}}/\widetilde{N}$ for $7\leq\widetilde{N}\leq100$ together with the bound of $k_{\mathrm{c}}/\widetilde{N}$ obtained in \eqref{eq:kc_bound}.
        We see that $k_{\mathrm{c}}/\widetilde{N}$ gets close to $K_{\mathrm{c}}$ as $\widetilde{N}\to\infty$.
    }
    \label{fig:kc}
\end{figure}

Assume that $k\geq\widetilde{N}K_{\mathrm{c}}-1/2+2\pi/(3\widetilde{N})$.
Then, from the mean value theorem and the monotonicity of $b(x)$, we have
\begin{align}
    t(K_{\mathrm{c}})-t\left(K_{\mathrm{c}}-\frac{1}{2\widetilde{N}}+\frac{2\pi}{3\widetilde{N}^{2}}\right)
    <\left(\frac{1}{2\widetilde{N}}-\frac{2\pi}{3\widetilde{N}^{2}}\right)b(K_{\mathrm{c}}).
\end{align}
Since $t(K_{\mathrm{c}})=0$ and $b(x)\leq 2$,
\begin{align}
    t\left(K_{\mathrm{c}}-\frac{1}{2\widetilde{N}}+\frac{2\pi}{3\widetilde{N}^{2}}\right)
    > -\frac{1}{2\widetilde{N}}b(K_{\mathrm{c}})+\frac{4\pi}{3\widetilde{N}^{2}}.
\end{align}
Hence we have
\begin{align}
    \frac{s_{k}}{\widetilde{N}}
    >& -\frac{1}{2\widetilde{N}}b(K_{\mathrm{c}})+\frac{4\pi}{3\widetilde{N}^{2}}
    +\frac{1}{2\widetilde{N}}b\left(\frac{k}{\widetilde{N}}\right)-\frac{\pi}{3N^{2}}\\
    >& -\frac{1}{2\widetilde{N}}\left[b(K_{\mathrm{c}})-b\left(K_{\mathrm{c}}-\frac{1}{2\widetilde{N}}+\frac{2\pi}{3\widetilde{N}^{2}}\right)\right]+\frac{\pi}{\widetilde{N}^{2}}.
\end{align}
Using the mean value theorem again gives
\begin{align}
    b(K_{\mathrm{c}})-b\left(K_{\mathrm{c}}-\frac{1}{2\widetilde{N}}+\frac{2\pi}{3\widetilde{N}^{2}}\right)=\left(\frac{1}{2\widetilde{N}}-\frac{2\pi}{3\widetilde{N}^{2}}\right)b'(x),
\end{align}
for some $x\in(K_{\mathrm{c}}-1/2\widetilde{N}+2\pi/(3\widetilde{N}^{2}),K_{\mathrm{c}})$.
Since $b'(x)$ is less than $4\pi$,
we obtain an evaluation of $s_{k}/\widetilde{N}$ as
\begin{align}
    \frac{s_{k}}{\widetilde{N}}&>-\frac{2\pi}{\widetilde{N}}\left(\frac{1}{2\widetilde{N}}-\frac{2\pi}{3\widetilde{N}^{2}}\right)
    +\frac{\pi}{\widetilde{N}^{2}}=\frac{4\pi^{2}}{3\widetilde{N}^{3}}>0,
\end{align}
meaning that $s_{k}>0$ as long as $k\geq\widetilde{N}K_{\mathrm{c}}-1/2+2\pi/(3\widetilde{N})$. From this, the desired evaluation holds:
\begin{align}
    k_{\mathrm{c}}\leq\left\lceil\widetilde{N}K_{\mathrm{c}}-\frac{1}{2}+\frac{2\pi}{3\widetilde{N}}\right\rceil
    \leq \widetilde{N}K_{\mathrm{c}}+\frac{1}{2}+\frac{2\pi}{3\widetilde{N}},
    \label{eq:kc_bound}
\end{align}
Figure~\ref{fig:kc} shows $k_{\mathrm{c}}/\widetilde{N}$ together with the derived bound.

\bibliography{dense}

\end{document}